# Single Stage Homogenous Deposition of Isolated Nano particles

H.Genish[1,a], N. Shabi[1] and M.Rosenbluh[1]

[1]*Department of Physics, Faculty of Exact Sciences, Bar-Ilan University, Ramat-Gan 5290002, Israel*

We demonstrate that a controlled and homogenous distribution of isolated nano-particles can be achieved in deposition by employing a single stage deposition technique ,utilizing a commercial ultrasonic humidifier to produce micrometer scale droplets from a suspension containing nano particles. Nano-diamonds were uniformly distributed on a Si substrate using this method. Particle density was controlled by varying the deposition time. The spatial distribution of the nano particles is shown to be uniform and with no aggregation. We propose that this method can be utilized in developing mass production processes of nano devices where a homogenous sub monolayer of nano-particles is required.

Nano particles have attracted much attention due to their potential application in various high-performance devices that make use of their unique electronic and optical properties[1–3]. In utilizing nano particles, one of the fundamental issues is how to deposit them uniformly onto a substrate with precise control of their density and without aggregation on the substrate. Many device examples can be found in the literature, for example, seeding a substrate for a subsequent CVD growth or preparation of bio sensors.

Various procedures have been developed in order to achieve a uniform monolayer of nano particles[4–6]. For instance Hong *et al*[7] have used the spin coating technique to obtain a sub monolayer of Co and Ag nano particles. However, like in most standard coating methods, their approach suffered from particle aggregation which occurred during the drying process. Recently Schmidlin *et al*[8] proposed using an electro-phoretic deposition technique which overcomes the clustering problem but requires that the substrate be conductive[9].

We demonstrate a simple method to overcome these limitations by which all types of nano-particles that can be suspended in a liquid can be deposited on virtually any substrate in a single step. The method is based on delivering the suspended nano particles to the substrate in small sub-micron diameter drops containing a single or small number of particles, while immediately evaporating the suspending fluid at the substrate surface.

Small droplets of the suspending fluid are created using an ultrasonic humidifier. The idea of using ultrasonic waves to produce small drops of fluids is a well established technique employed in spray coating and is used to produce scalable thin films. Ultrasonic assisted spray systems are also used in the preparations of the nano particles themselves[10–12] in processes such as spray pyrolysis. However, unlike spray pyrolysis in which the raw material for forming the nano particlesis delivered by the drops we suggest using the droplets as carriers for the pre-fabricated nano particles in a manner that enable the deposition of isolated particles. Thus any type of suspendible particle may be deposited using our method, irrespective of the process required to produce the particle itself.

The deposition system we used consists of a standard household ultra-sonic humidifier (SC-609, Procare) a plastic container and a hot-plate. The ultra-sonic humidifier uses a piezoelectric transducer to create high frequency mechanical oscillations in a metal diaphragm. These oscillations form a fine mist of droplets at micrometer scale[13,14]. The droplets have the same composition as the original suspension and thus carry with them the suspended nano particles. The droplets are

---

[a] Corresponding author, Email : hadar.genish@biu.ac.il

ejected from the humidifier into a container which confines the vapor and prevents air turbulence from the surrounding environment. A hot plate is used to increase the temperature of the substrate in order to rapidly evaporate the suspending liquid of a droplet which lands on the substrate, thus inhibiting nano particle agglomeration

Provided that the nano particle concentration in the suspension is low enough so that each drop contain a single or small number of nano particle, the film formed by this method should be a homogenous layer of single isolated nano particles. The size of the drop sets the limit of the maximum concentration of nano particles in the original suspension. The device used in our experiment operates at a fixed frequency and thus lacks the option of control of the drop size or diameter distribution.. The droplet diameter was estimated to be 3 micrometer from the ring pattern of nano-particles left on the substrate when using suspensions with a high concentration of particles.

To demonstrate the technique we used a nano-diamond(ND) slurry (MSY-0.05GAF, Microdiamant) suspended in DI water at a ratio of 1:20.Although the original slurry contained mostly isolated ND, clusters of them are still present and we removed these from the suspension by ultracentrifuging the liquid suspension at 25,000 g .The supernatant was then diluted again and a series of depositions with various concentrations of nano diamonds were performed in order to determine the concentration for which no agglomeration could be seen. The NDs were deposited on Si and SiO2/Si wafers of 15x15 mm size for various durations in order to determine the deposition rate of our experimental configuration.

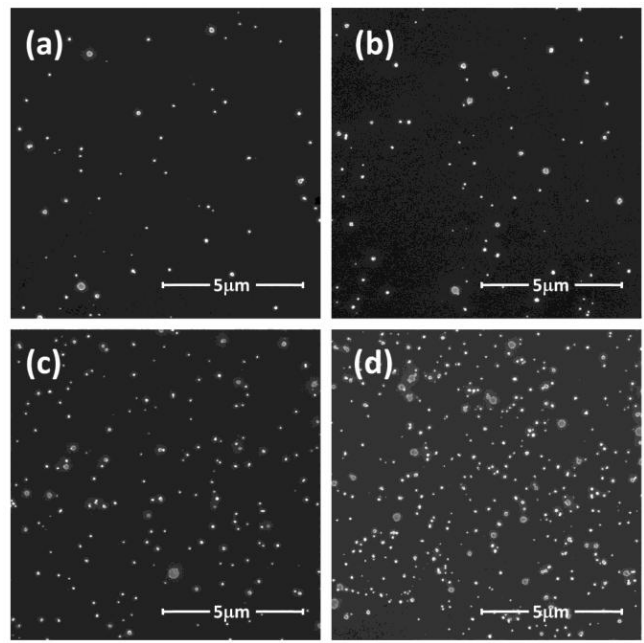

**Figure 1 : SEM micrograph of nano diamond in a sub-monolayer deposited for (a) 30min (b) 60 min (c) 120 min and (d) 180 min using a supernatant from 10,000g ultracentrifuged slurry diluted in DI water at a ratio of 1:15.**

Figure 1 shows SEM micrographs of the deposited NDs for various durations. From Fig 2 (a) it can be seen that the deposition was homogenous on a larger scale this was verified by an analysis of different zones on each sample.

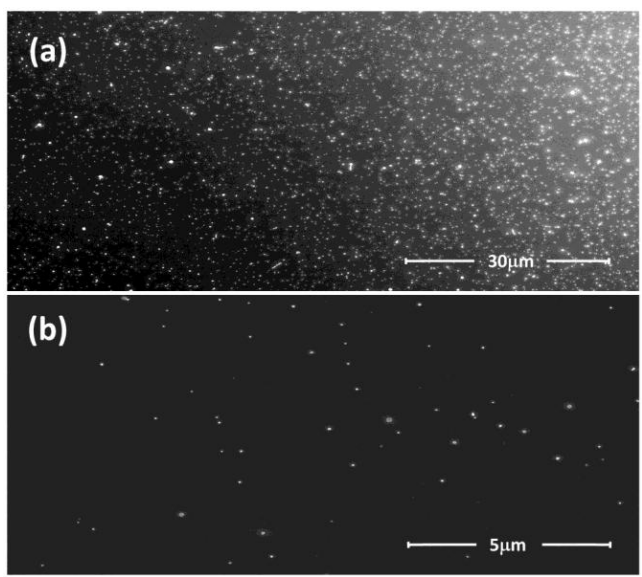

**Figure 2 : SEM micrographs of the (a) a large area deposited with nano particle showing the homogeneity of the deposition. (b) Zoom in showing the individual particles.**

Figure 3 shows the NDs density as a function of deposition time as analyzed from SEM micrographs using ImageJ[15]. The deposition rate was calculated to be 0.1 particles/$\mu m^2$ per minute of deposition. It is worth noting that even for long deposition times the coverage remains uniform.

Fig 4 shows an atomic force microscope (AFM) micrograph of the deposited sample. The AFM results were analyzed to confirm the low probability of aggregates as can be seen in Fig 5.

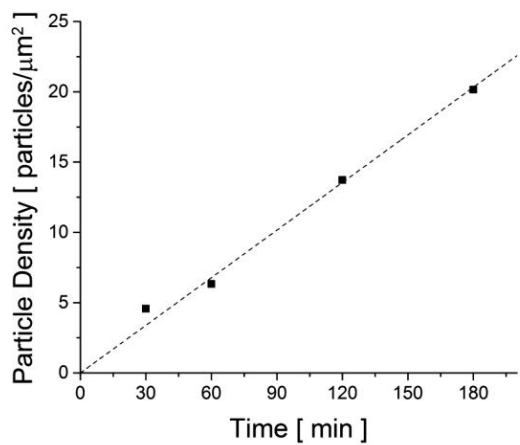

**Figure 3 : Particles density as a function of deposition time**

The results of our study suggest that ultrasonic assisted spray coating can be applied to nano particle deposition and provides a simple and low cost method to deposit isolated nano particles without the need for chemical modification. The technique was demonstrated using ND and may be used for applications such as the growth of thin diamond layers through CVD or the development of biocompatible sensors. Scaling the process can be achieved using an ultrasonic nozzle rather than a membrane to remove limitations on the dimension of substrate as well as increasing the efficiency of the process in terms of material loss.

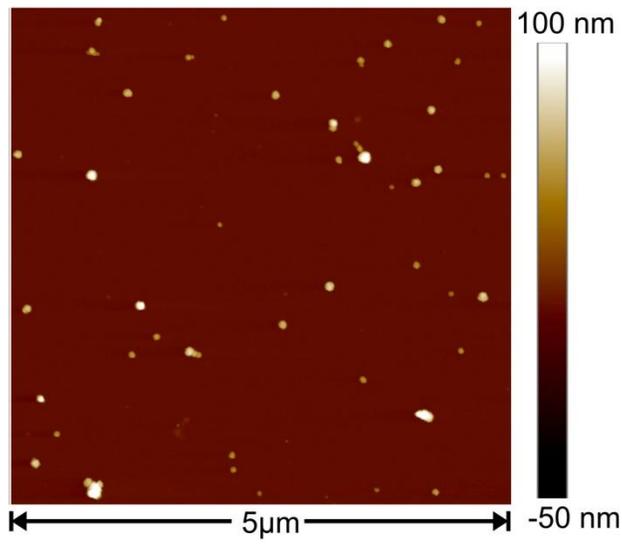

**Figure 4 : Typical AFM micrograph of the deposited samples**

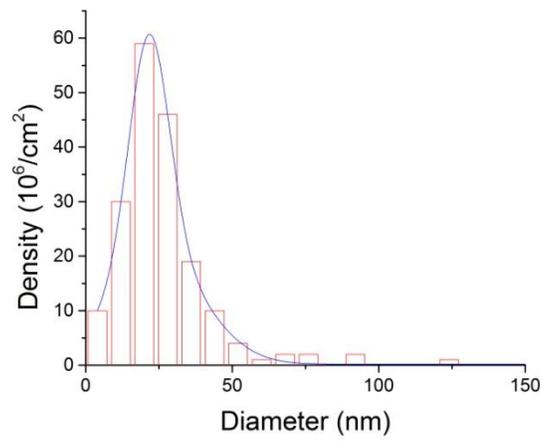

**Figure 5 : Particle analysis of the AFM micrographs from the deposited sample.**